\documentclass[11pt]{article}
\usepackage{isolatin1,latexsym,graphicx,a4wide,cite}
\usepackage{subfigure}

\title{New insights into pedestrian flow through bottlenecks}

\author{Armin Seyfried$^1$, Tobias Rupprecht$^2$, Oliver Passon$^1$,\\ Bernhard Steffen$^1$, Wolfram Klingsch$^2$ and Maik Boltes$^1$\\ [0.2cm] 
\footnotesize $^1$J\"ulich Supercomputing Centre, Research Centre J\"ulich, 52425 J\"ulich, Germany\\
[-0.1cm]\footnotesize $^2$Institute for Building Material Technology and Fire Safety Science, University of Wuppertal\\ 
[-0.15cm]\footnotesize Pauluskirchstrasse 11, 42285 Wuppertal, Germany\\
\footnotesize  E-mail: a.seyfried@fz-juelich.de, rupprecht@uni-wuppertal.de,
o.passon@fz-juelich.de,\\
[-0.15cm]\footnotesize b.steffen@fz-juelich.de, klingsch@uni-wuppertal.de, m.boltes@fz-juelich.de\\
}

\date{\footnotesize 18.10.2007}

\begin{document}

\maketitle

\begin{abstract}
\par 
\noindent
Capacity estimation is an important tool for the design and dimensioning of 
pedestrian facilities. The literature contains different procedures and 
specifications which show considerable differences with respect to the
estimated flow values. Moreover do new experimental data indicate a 
stepwise growing of the capacity with the width and thus challenge the validity 
of the specific flow concept. To resolve these differences we have studied 
experimentally the unidirectional pedestrian flow through bottlenecks under 
laboratory conditions. The time development of quantities like individual 
velocities, density and individual time gaps in bottlenecks of different 
width is presented. The data show a linear growth of the flow with the width. 
The comparison of the results with experimental data of other authors indicates 
that the basic assumption of the capacity estimation for bottlenecks has to be 
revised. In contradiction with most planning guidelines our main result is, that 
a jam occurs even if the incoming flow does not overstep the capacity defined 
by the maximum of the flow according to the fundamental diagram.
\end{abstract}

\section{Introduction}
\label{SEC-INTRO}

Today more than half of the mankind is living in cities, while 50 years ago this 
were only 30 percent \cite{UN05}. The growth of cities worldwide is inexorable 
and a great challenge for the urban development and architecture. In this context 
knowledge about pedestrian dynamics is important and allows e.g. the design and 
optimization of facilities with respect to safety, level of service and economy. 
A basic practical application is the capacity estimation which allows the 
dimensioning or evaluation of facilities 
\cite{PRED78ENG,FRUIN71,HCPM85,WEID93,SFPE02}. For this purpose one is interested 
in the number of pedestrians, $\Delta N$, which is able to pass the facility in a 
certain time interval, $\Delta t$. The specifying quantity is the flow: 

\begin{equation}
J = \frac{\Delta N}{\Delta t} \,.
\label{DNDT}
\end{equation}

The capacity is defined as the maximal value for the flow, $C=J_{max}$, and allows 
e.g. the estimation of the minimal time needed for emptying of a given facility or 
of the minimal width of the facility which allows the emptying in a given time. In 
the case of a classical bottleneck, like a narrowing in a corridor, it is generally 
assumed that a jam will occur when the incoming flow exceeds the capacity of the 
bottleneck \cite{PRED78ENG,WEID93,NELS02,helbing06a}. At first glance this seems to 
be a reasonable assumption which can be justified by reference to the continuity 
equation. This equation trivially implies the increase of the density in a control 
volume if the incoming flow is larger than the outgoing flow. However, this 
establishes $J_{max}$ only as an upper limit for the incoming flow which produces a 
jam and does not exclude its formation for smaller flow values already.   

In the literature one can find different specifications to estimate the capacity of 
a pedestrian facility. The flow equation in combination with empirical measurements 
is commonly used, see for example 
\cite{OLD68,NAV69,PRED78ENG,CAR70,FRUIN71,PUSH75,HCPM85,WEID93,NELS02}. 
With the width of the pedestrian facility, $b$, the flow equation can be written 

\begin{equation}
J = \underbrace{\rho\;v}_{=J_s}  b \,. 
\label{BASFLOW}
\end{equation}

The specific flow, $J_s$, gives the flow per unit-width and can be calculated as the 
product of the average density, $\rho$, and the average speed, $v$, of a pedestrian 
stream. The velocity and thus the flow is a function of the density. The empirical 
relation between flow and density, $J=J(\rho)$, is called the fundamental diagram 
for which one can find different representations. With a given fundamental diagram 
in the representation $J(\rho)$ the capacity of a facility is defined as the maximum 
of this function. In general it is assumed that for a given facility (e.g. corridors, 
stairs, doors) Equation \ref{BASFLOW} is valid, which implies that the fundamental 
diagrams for different $b$ merge into one universal diagram for the specific flow 
$J_s$. Consequently the capacity, $C$, is assumed to be a linear function of the 
width, $b$. This assumption is used for all kinds of facilities and pedestrian streams 
(uni- or bidirectional). In this article we concentrate on unidirectional pedestrian 
movement through bottlenecks under normal conditions. If in the following the term 
movement under normal conditions is used it is meant, that panic or in particular 
non adaptive behavior which can occur in critical situation or under circumstances 
including rewards \cite{MINT51} are excluded.

\begin{figure}[thb]
\begin{center}
 \includegraphics[width=.65\textwidth]{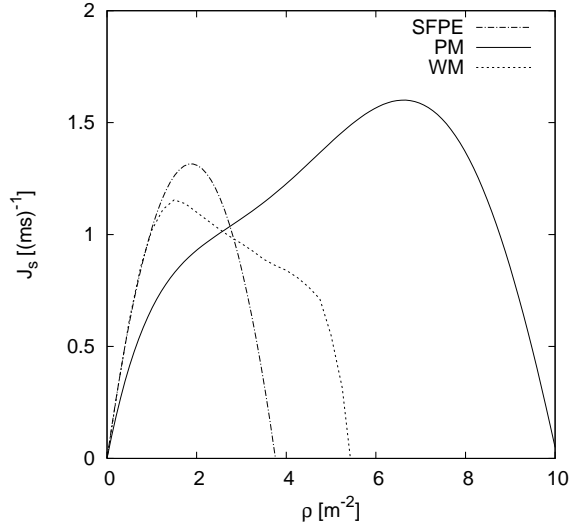}
\caption{Fundamental diagrams for pedestrian movement through openings and doors 
according to the SFPE Handbook SFPE \cite{NELS02} and the guidelines of Predtechenskii 
and Milinskii PM \cite{PRED78ENG} and Weidmann WM \cite{WEID93}. \label{R-FS-INTRO}}
  \end{center}
\end{figure}

Figure \ref{R-FS-INTRO} shows various fundamental diagrams used in handbooks and 
planing guidelines for the capacity estimations of openings, doors or bottlenecks. The 
fundamental diagrams differ with respect to both, the height of the maximum (i.e. the 
capacity) and the corresponding density value. According to the handbook of the Society 
of Fire Protection Engineers (SFPE) and the guideline of Weidmann (WM) the maximum 
is located at a density around $\rho=1.8\; m^{-2}$, while the maximum considered from
Predtechenskii and Milinskii\footnote{We note, that the data of PM are gained 
from Equation 7, 8 and 9 in chapter III in \cite{PRED78ENG} and not from the tabular in 
the appendix of the book. For the scaling of the density we chose $A=0.113\;m^2$ which 
gives the area of horizontal projection of a pedestrian in mid-season street dress and 
the fundamental diagram for movement under normal conditions.} (PM) is located at 
densities larger than $\rho=7\;m^{-2}$. The predicted capacities differ from 
$C=1.2\;(ms)^{-1}$ to $1.6\;(ms)^{-1}$. But as mentioned above all this procedures 
assume a linear dependence between capacity and bottleneck-width.
 
Contrary to this, Hoogendoorn and Daamen \cite{Hoogendoorn03a,Hoogendoorn05a} claim 
that the capacity is growing in a step-wise manner. This statement is based on their 
observation that inside a bottleneck the formation of lanes occurs, resulting from the 
zipper effect during entering the bottleneck. The zipper effect is a self organization 
phenomenon leading to an optimization of the available space and velocity inside the 
bottleneck. The data in \cite{Hoogendoorn03a,Hoogendoorn05a} indicate that the 
distance between these lanes is independent of the bottleneck-width. This would imply 
that the capacity increases only when an additional lane can develop, i.e. that this 
would occur in a stepwise manner with increasing width \cite{Hoogendoorn05a}. 
Consequently, either the specific flow would decreases between the values where the 
steps occur or the flow equation in combination with the concept of a specific flow 
would not hold. One goal of this work is to examine this claim (and especially 
Section~\ref{RESULTS} is devoted to this question). 

\begin{table}[h]
\begin{center}
\begin{tabular}{| c || c | c | c | c || c | c | c | c|}
\hline
\multicolumn{1}{| c ||}{$b\;[m]$} & \multicolumn{7}{| c |}{\textbf{$C=J_{max} [s^{-1}]$}} \\
\hline
\multicolumn{1}{| c ||}{} & \multicolumn{4}{| c ||}{\textbf{Estimation}} & \multicolumn{3}{| c |}{\textbf{Measurement}} \\
\hline
    & SFPE & PM   & WM   & HG & Kretz & Nagai & M\"uller  \\ \hline
\hline                                    
0.8 & 1.04 & 1.28 & 0.98 &      & 1.43  & 2.58  &           \\ 
0.9 & 1.17 & 1.44 & 1.10 &      & 1.58  &       & 2.26      \\ 
1.0 & 1.30 & 1.60 & 1.23 & 0.78 & 1.85  &       &           \\ 
1.2 & 1.56 & 1.92 & 1.47 & 0.78 & 2.15  & 3.27  & 2.75      \\ 
1.5 & 1.95 & 2.40 & 1.84 & 1.56 &       &       & 3.59      \\ 
1.6 & 2.08 & 2.56 & 1.96 & 1.56 &       & 3.86  &           \\ \hline
\end{tabular}
\end{center}
\caption[dummy]{Comparison of estimated and measured capacities for bottlenecks.
The results reveal the differences not only between measurements and estimations, 
but also among the estimation methods or specifications.  
\label{CAPEST}}
\end{table}

In Table \ref{CAPEST} these uncertainties in the capacity estimations are compared 
with selected experimental data on pedestrian flow. The table includes the methods 
according the SFPE Handbook \cite{NELS02}, the guideline of Predtechenskii and 
Milinskii (PM) \cite{PRED78ENG}, the guideline of Weidmann (WM) \cite{WEID93}, the 
specification of Hoogendoorn (HG) in \cite{Hoogendoorn05a} and the experimental 
measurements by Kretz \cite{KRETZ06c}, Nagai \cite{NAGAI06} and M\"uller 
\cite{Mueller81}. 
Some authors introduce an effective width, taking into account that pedestrians keep 
a distance to walls. The clear width, measured from wall to wall, is reduced in 
dependence of handrails, wall structure or obstacles.     
For a better comparability we neglected the effective width concept used in 
\cite{NELS02,WEID93} as well as in \cite{Hoogendoorn05a}. To consider the 
effective width would lead to smaller flows and larger differences to the experimental 
measurements. Table \ref{CAPEST} reveals the differences among the results of 
estimation procedures, which can exceed a factor of two. Given that these estimations 
come without error margins there is, strictly speaking, no contradiction. However in 
the context of evacuation-time predictions a discrepancy of 100 percent is 
unacceptable. The experimental measurements of the capacity of bottlenecks show also 
non negligible variations. The variation between experimental data and estimation 
procedures can exceed a factor of four, see Table \ref{CAPEST}. For the differences 
among the experimental 
measurements there are multiple possible reasons. However one has to note that all 
measurements were performed under well controlled laboratory conditions and that in 
all experiments the test persons were asked to move normally. Hence the influence 
of panic or pushing can be excluded. Other possible reasons are e.g. the geometry of 
the bottleneck or the initial conditions in front of the bottleneck. A detailed 
discussion will be possible together with the experimental results presented in this 
paper and can be found in Section \ref{RESULTS}.

Another important question concerns the situation when a jam in front of a facility 
is present. Predtechenskii and Milinskii \cite{PRED78ENG} assume that in this case 
the flow through the bottleneck is determined by the flow in front of the bottleneck. 
They suppose that the density inside the jam will be higher than the density 
associated with the capacity and thus it is possible that the reduced flow in front 
of the bottleneck will cause a flow through the bottleneck smaller than the 
bottleneck-capacity.

Recapitulating the above discussion there are several open questions. The central 
one is if the flow equation in combination with the concept of a specific flow is 
adequate for capacity estimations leading to a linear increase of the capacity with 
the width. The next question is why the experimental flow measurements exceed the 
estimated capacities up to a factor of four. The closing question is which flow will 
tune in if a jam occurs in front of the bottleneck and if it is conserved at the value 
of the capacity. 

To answer this questions and to resolve the discrepancies among the estimation 
results an experiment is arranged where the density and the velocity and thus the 
flow inside the bottleneck is measured while a jam occurs in front of the bottleneck. 
Our experiment is performed under laboratory conditions with a homogeneous group of 
test persons and equal initial conditions for the density and position of the test 
persons in front of the bottleneck. Exclusively the width of the bottleneck and the 
number of the pedestrians are varied. For the analysis the trajectories inside the 
bottleneck are determined and used to resolve the time dependence of the flow, the 
density and the velocity.

\section{Experimental setup}

\begin{figure}[htb]
\begin{center}
 \includegraphics[width=.27\textwidth]{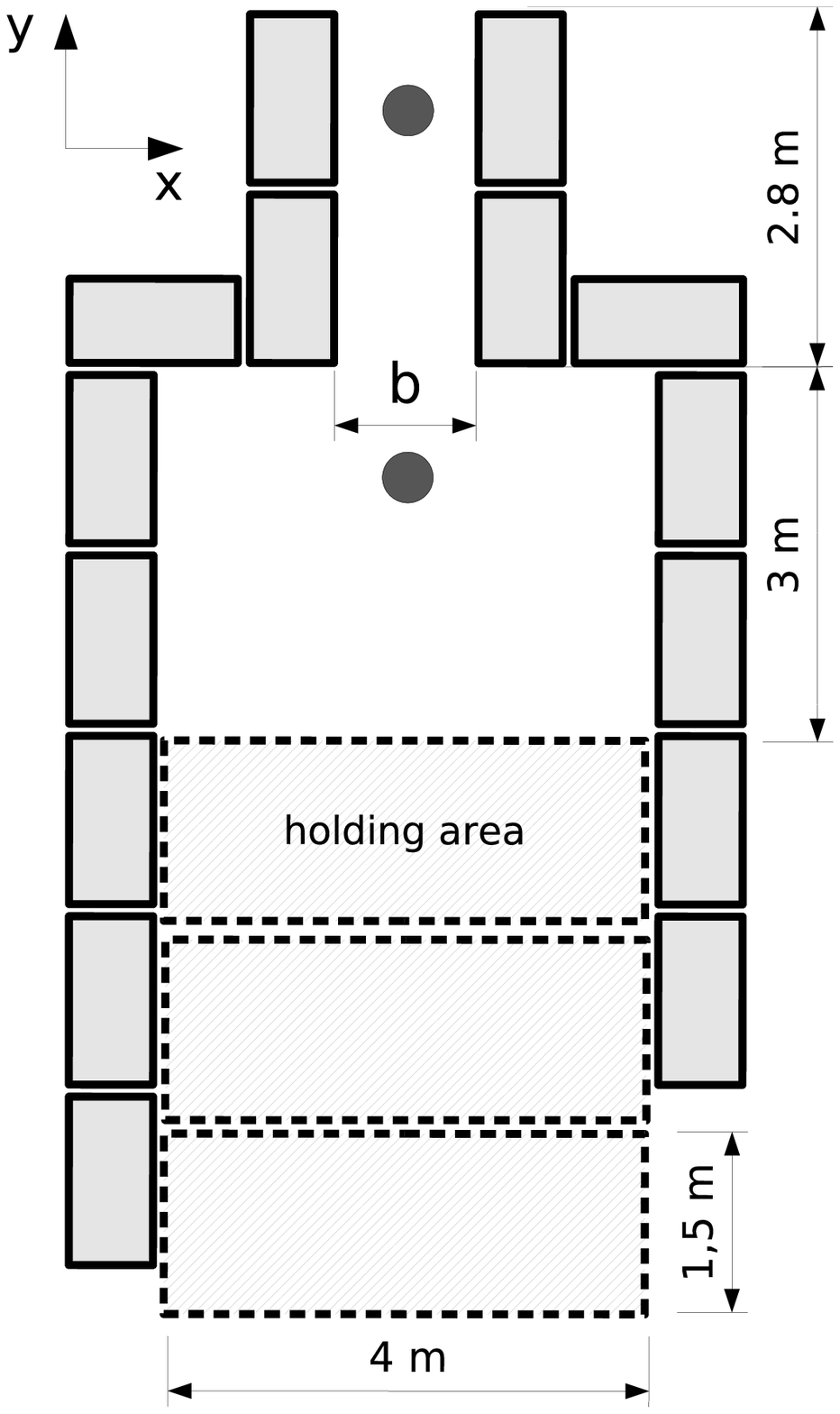}
\hspace{20.pt}
 \includegraphics[width=.39\textwidth]{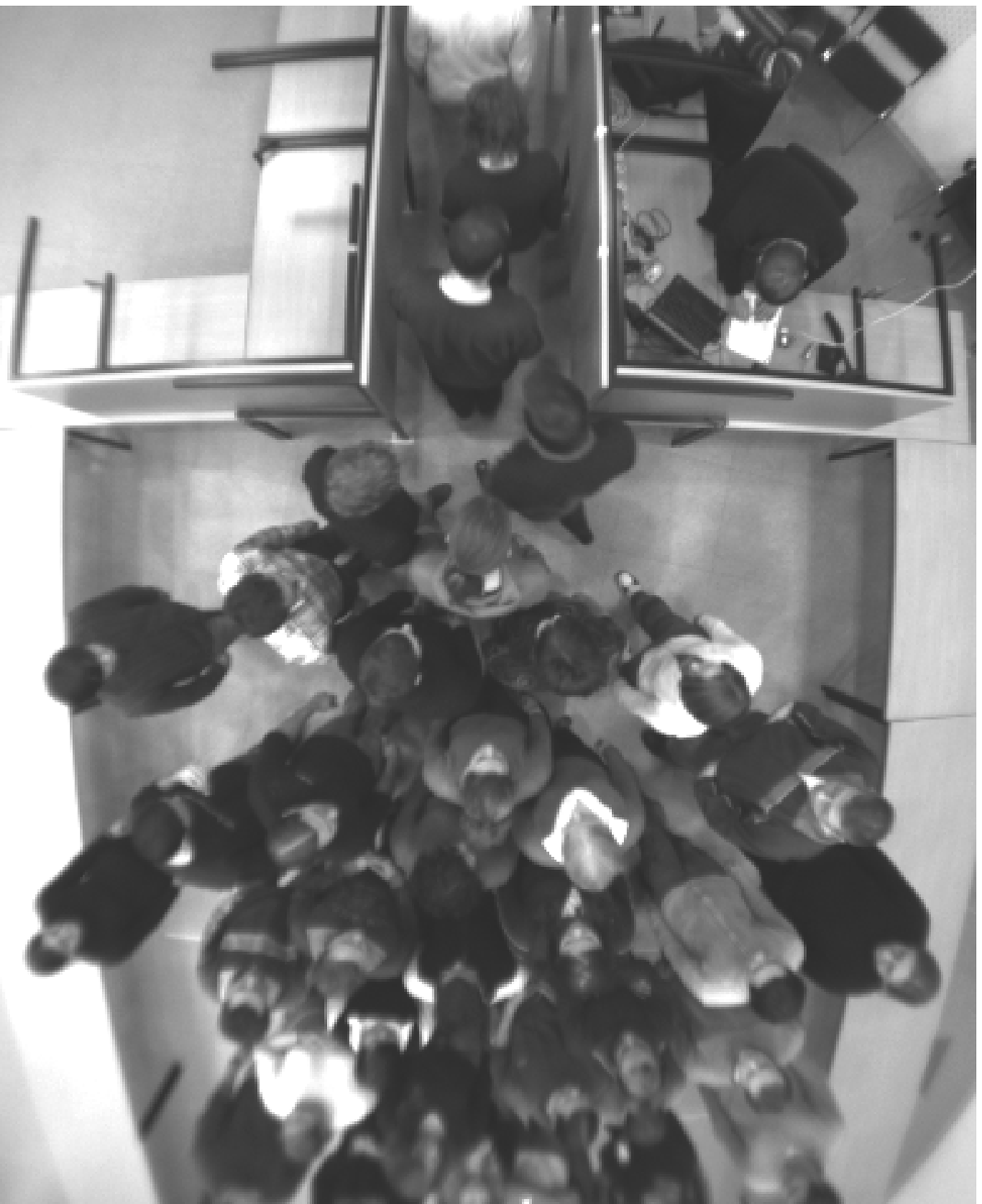}
\caption{Experimental setup. In the left drawing the position of the video cameras 
are marked with circles. The holding and measurement areas are hatched. The  
photo shows the situation for b=0.8 m in front of the entrance in the bottleneck.
  \label{ESETUP}}
  \end{center}
\end{figure}

For the experimental setup the given priority is to measure the influence of the 
bottleneck width and to exclude as far as possible other effects which can influence 
the flow. Thus one main point was to ensure same conditions of all runs with different 
widths. Moreover one has to consider the limited resources and thus to study the 
dependence of the results on the number of test persons.
Our experiment was arranged in the auditorium `Rotunde' at the Central Institute for 
Applied Mathematics (ZAM) of the Research Centre J\"ulich. The configuration is 
shown in Figure \ref{ESETUP}. The group of test persons was composed of 
students and ZAM staff. The boundary of the corridor in front of the bottleneck 
and the bottleneck was arranged from desks. The height of the bottleneck assured a 
constant width from the hips to the shoulders of the test persons. The length of the 
bottleneck amounted to $l_{bck}=2.8\;m$. The holding areas ensured an equal initial 
density of the pedestrian bulk in front of the bottleneck for each run. The distance 
from the center of the first holding area to the entrance of the bottleneck was three 
meter.

A stepwise increase of the flow due to lane formation is expected more pronounced 
for small numbers of lanes. Bottleneck widths smaller than $b=0.8 \; m$ are not 
relevant in practice. For $b>1.2 \; m$ it was estimated that even with $N=60$ a 
steady state will hardly be reached.
Thus the width of the bottleneck was increased from the minimal value of 
$b=0.8 \; m$ in steps of $0.1\;m$ to a maximal value of $b=1.2 \; m$. For every 
width we performed three runs with $N=20$, $40$ and $60$ pedestrians in front of 
the bottleneck. All together 18 runs were performed to analyze the effect of the 
bottleneck-width, $b$, and the influence of the number of pedestrians, $N$. At 
the beginning of each run $N$ test persons were placed in the holding areas with 
a density of $\rho_{ini}=3.3\;m^{-2}$. The test persons were advised to move 
through the bottleneck without haste but purposeful. It was emphasized not to 
push and to walk with normal velocity. The test persons started to move after an 
acoustic signal. The whole cycle of each run was filmed by two cameras, one 
situated above the center of the bottleneck and the other above the entrance 
of the bottleneck (indicated by the circles in Figure \ref{ESETUP}). 

\section{Data analysis}
\label{ANA}

\subsection{Specific flow}

In a first step we have calculated the flow on the basis of the crossing time of the 
first and the last pedestrian, according Equation \ref{DNDT}. 

\begin{table}[htb]
\begin{center}
\begin{tabular}{| c || c | c | c |}
\hline
\multicolumn{1}{| c ||}{$b\;[m]$} & \multicolumn{3}{| c |}{$J_s\;[(ms)^{-1}]$} \\
\hline
    & $\Delta N = 60$ & $\Delta N = 40$ & $\Delta N = 20$ \\ \hline
\hline
0.8 & 1.61 & 1.77 & 1.86 \\
0.9 & 1.86 & 1.91 & 2.06 \\
1.0 & 1.90 & 2.08 & 2.19 \\
1.1 & 1.93 & 1.93 & 1.78 \\
1.2 & 1.97 & 1.81 & 2.31 \\ \hline
\end{tabular}
\end{center}
\caption[dummy]{ The specific flow, $J_s$, is calculated by $\Delta N / (\Delta t\;b)$. 
The time interval $\Delta t$ between the passage of the first person and last person 
is measured at the center of the bottleneck at $y=0.4\; m$. 
 \label{NAVPHI}}
\end{table}

In Table \ref{NAVPHI} we collect these values divided by the width, $b$, to examine 
if they approach a constant value for the specific flow, $J_s$. For $N=60$ the data 
show a small but systematic increase of specific flow with width indicating an 
influence of the zipper effect and of the boundaries of the bottleneck, where the 
optimization through the zipper effect can not act. Nevertheless this variation is 
small compared to the variation in Table \ref{CAPEST} and for $N < 60$ partly washed 
out by fluctuations. Furthermore for small $N$ the time is not sufficient to reach a 
stationary state. To control the influence of fluctuations, non stationary states and 
how the zipper effect acts, the trajectories and the time development of density 
$\rho(t)$, velocity $v(t)$ and flow $J(t)$ are analyzed. 

\subsection{Trajectories and probability distributions \label{hoogi_macht_mist}}

\begin{figure}[hbt]
\begin{center}
 \includegraphics[width=.4\textwidth]{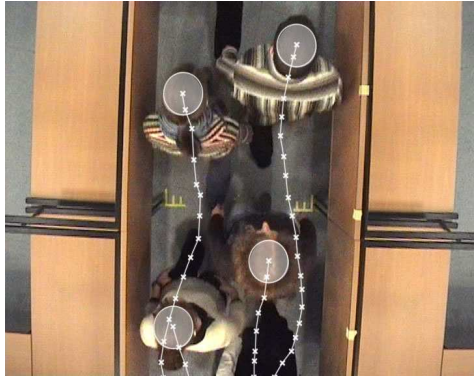}
\caption{A snapshot from the camera above the bottleneck. The trajectory are 
determined by marking the center of the head of each person manually.
  \label{TRAJ}}
  \end{center}
\end{figure}

For the determination of the trajectories a manual procedure based on the standard 
video recordings of a camera above the bottleneck is used. The camera recording 
is done in digital CCIR 601 format on Mini-DV tape: PAL size of $720 \times 576$ 
rectangular pixel with aspect ratio of $1.0666:1$ and a frame rate of 25 frames per 
second. These recordings are analyzed with the software tool 
{\it Adobe After Effects} \cite{ADOBE}. To determine the trajectory the center of 
the head is marked and followed in time manually. To facilitate the positioning of 
the marker on the center of the head by hands and by visual judgment a circle around 
the marker is drawn, see Figure \ref{TRAJ}. For this procedure every second frame 
(80 ms) is used. 

The trajectory of each pedestrian is exported to the format 
{\it Adobe Motion Exchange} and re-processed. The re-processing is used to 
minimize the errors by checking the plausibility of the data and correcting 
distortions based on pixel sizes and camera perspective. The plausibility 
check looks for backward movement, speed limits, and positions of singular 
trajectory points in between the boundaries. To ensure equally scaled 
measurements in pixel row and column the correction of the distortion has 
to scale the pixels to quadratic ones. Furthermore the unit meter per pixel 
is measured at several reference points to scale pixel sizes to real metric 
values. Then the picture is rectified because of a slightly non-perpendicular 
camera view. No radial distortion is required. The last step rotates and moves 
the coordinate system to the left bottom corner of the bottleneck with the 
negative y axis in direction of movement.

For the further data analysis the individual trajectories $(x_{i,j},y_{i,j},t_{i,j})$ 
were used. The index $i$ marks the pedestrian, while $j$ marks the sequence of
the points in time. To study the microscopic properties and the time dependence of the 
flow the individual time gap, $\Delta t_i$, is introduced

\begin{equation}
J = \frac{\Delta N}{\Delta t} = \frac{1}{<\Delta t_i>} \quad \mbox{with} \quad 
<\Delta t_i>=\frac{1}{N-1}\sum_{i=1}^{N-1}t_{i+1}-t_{i}.
\label{ITG}
\end{equation}

Where $<\Delta t_i>$ is the mean value of the time gaps between the crossings of two 
following pedestrians. 

\begin{figure}[th]
\begin{center}
  \includegraphics[width=0.62\textwidth]{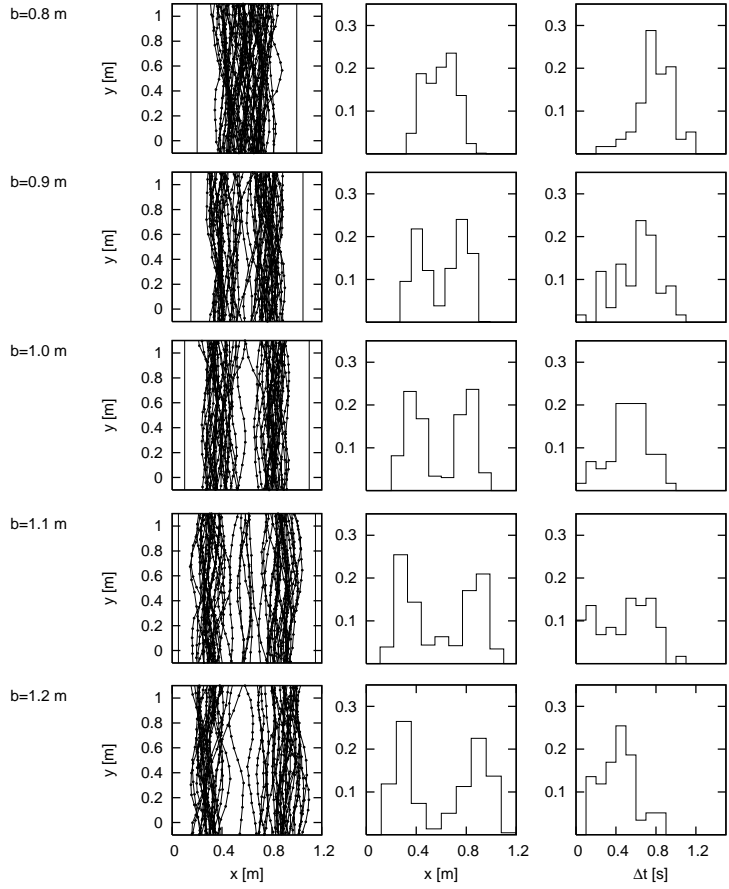}
\caption{For the runs with $N=60$ and from top to bottom with increasing $b$: The 
trajectories (left), the probability to find a pedestrian at position $x$ (middle) 
and probability distribution of the time gaps $\Delta t_i$ at $y=0.4\;m$ (right). 
For $b \geq 0.9 \;m$ the formation of lanes is observable. However the distance 
between the lanes increases continously with $b$ leading to a continuous decrease 
of time gaps between two following pedestrians, see Figure \ref{ZIPPER}. Thus no 
indications of a stepwise change of the flow can be found. 
  \label{MULTITRJ}}
  \end{center}
\end{figure}

In Figure \ref{MULTITRJ} we have collected for the runs with $N=60$ the trajectories, 
the probability distribution to find a pedestrian at the position $x$ averaged over 
$y$ and the probability distribution of the individual time gaps, $\Delta t_i$, at 
center of the bottleneck at $y=0.4\;m$. For the determination of the time gaps at 
$y = 0.4\;m$ we interpolate linearly between adjacent trajectory points with the 
current velocity. The double peak structure in the probability distribution for 
$b \geq 0.9 \;m$ of the positions display the formation of lanes. The separation of 
the lanes is continuously growing with the width of the bottleneck. As a consequence 
of the zipper effect one expects also a double peak distribution for the time gaps 
$\Delta t$. However this is not so articulated as in the separation of lanes in the 
position. One can only observe a broadening of the time gap distribution with 
increasing $b$ and a drift to smaller values. The changes in these figures as a 
function of the width are all continuous except for the transition from one to two 
lanes and there are no indications of a stepwise increase or decrease in any 
observable. 

\begin{figure}[thb]
\begin{center}
 \includegraphics[width=.5\textwidth]{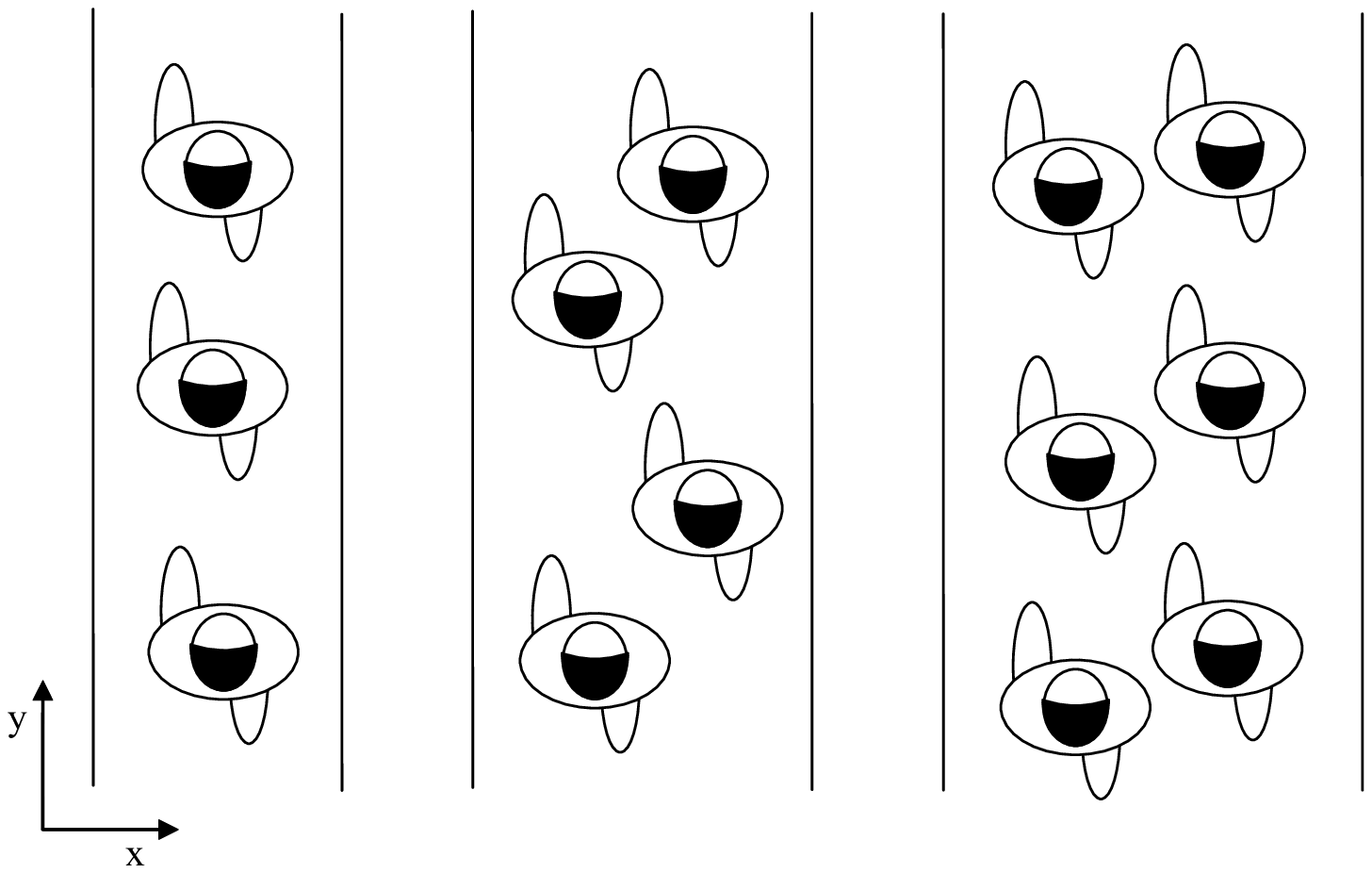}
\caption{A sketch of the zipper effect with continuously increasing lane distances 
in $x$: The distance in $y$, the walking direction, decreases with increasing distance 
in $x$. This leads to a monotonous increase of the flow with increasing width.}
  \label{ZIPPER}
  \end{center}
\end{figure}

\subsection{Time dependence of $\rho$, $v_i$, and $\Delta t_i$}
\label{TIMDEV}

For the first pedestrian passing the bottleneck in a run the velocity and density 
will be different from the velocity and the density of the following pedestrians. 
One expects that the density will increase while the velocity will decrease in time.
A systematic drift to a stationary state, where only fluctuation around a constant 
value will occur, is expected. For the analysis of the time dependence of the 
individual velocities, $v_i$, and density, $\rho$, we calculate the following 
quantities 

\begin{equation}
 v_i(t_{i,j}) = \frac{y_{i,j+2}-y_{i,j-2}}{t_{i,j+2}-t_{i,j-2}}\,. 
\label{VEL} 
\end{equation}

The time, $t_{i,j}$, corresponds to the first trajectory point 
$(x_{i,j},y_{i,j},t_{i,j})$ behind $y=0.4\;m$, the center of the bottleneck. For the 
density $\rho(t)$ we use the momentary number $n(t)$ of persons inside the measurement 
section reaching from $y=[0.9\;m,-0.1\;m]$ at the instant $t$, see Figure \ref{TRAJ}. 
With the unit length of the observation area the density is 
  
\begin{equation}
 \rho(t)=\frac{n(t)}{b}\,.
\label{DEN} 
\end{equation}

\begin{figure}[thb]
\begin{center}
 \includegraphics[width=.7\textwidth]{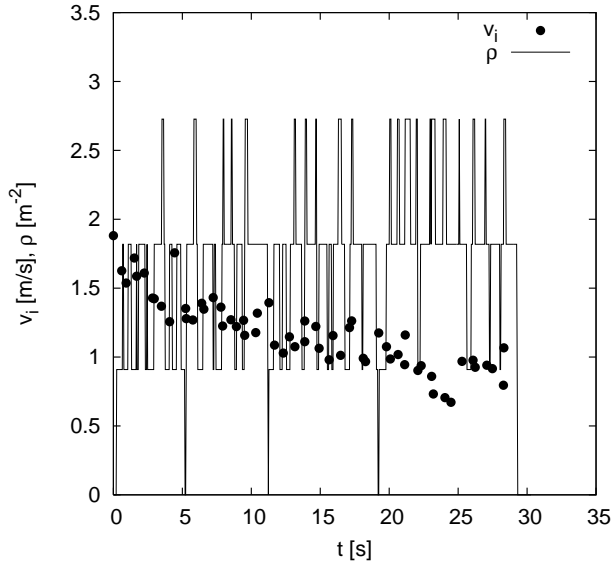}
\caption{Development of individual velocity, Eq. \ref{VEL}, and density, Eq. \ref{DEN}, 
in time for the run with $N=60$ and $b=1.1\;m$. While the velocity decreases the 
density increases.}
  \label{V-R-TIME}
  \end{center}
\end{figure}

Figure \ref{V-R-TIME} shows the time-development of the individual velocities and the 
density for the run with $N=60$ and $b=1.1\;m$. Plots for other runs can be found in 
\cite{RUPDIP06}. The concept of a momentary density in this small observation area 
is problematic because of the small (1-4) number of persons involved and leads to large 
fluctuations in the density. But one can clearly identify the decrease of the velocity 
and the increase of the density. 

\begin{figure}[h]
\begin{center}
 \includegraphics[width=.7\textwidth]{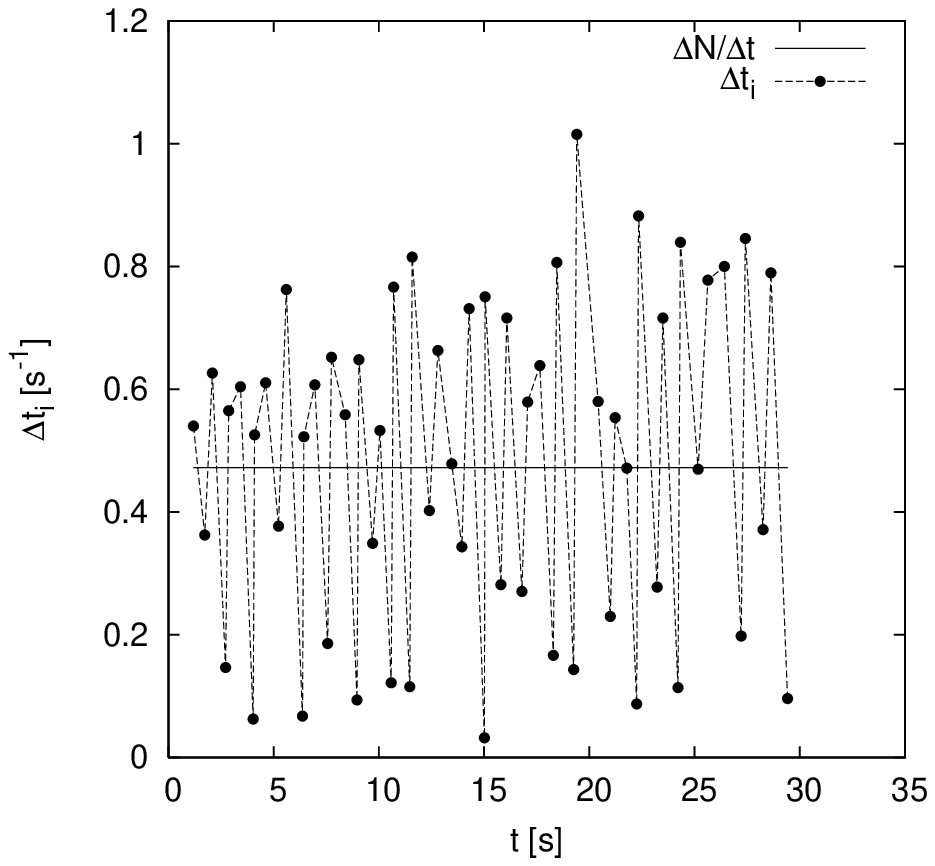}
\caption{Development of the individual time gaps for the run with $b=1.1\;m$ and 
$N=60$ in time. The zipper effect is responsible for the systematic alternations 
from small to large time gaps.}
  \label{TG-TIME}
  \end{center}
\end{figure}

For the individual time gaps, see Eq.~\ref{ITG}, a time dependence or a trend to a 
stationary state is hard to identify, see Figure \ref{TG-TIME}, because the velocity 
decrease and the density increase compensate largely. A potential time dependence is 
covered by large and regular jumps from small to high time gaps caused by the zipper 
effect. 

To find stationary values for the velocity and density by means of regression analysis 
the tool {\em MINUIT} \cite{MINUIT} for function minimization is used with the 
following model function borrowed from relaxation processes:

\begin{equation}
 f(t)  =  f_{stat} + A \exp{-\frac{t}{\tau}} \quad \mbox{for} \quad f(t)=v(t) \quad 
\mbox{and} \quad f(t)=\rho(t)
\label{FITEQ} 
\end{equation}

\begin{figure}[h]
\begin{center}
 \includegraphics[width=.7\textwidth]{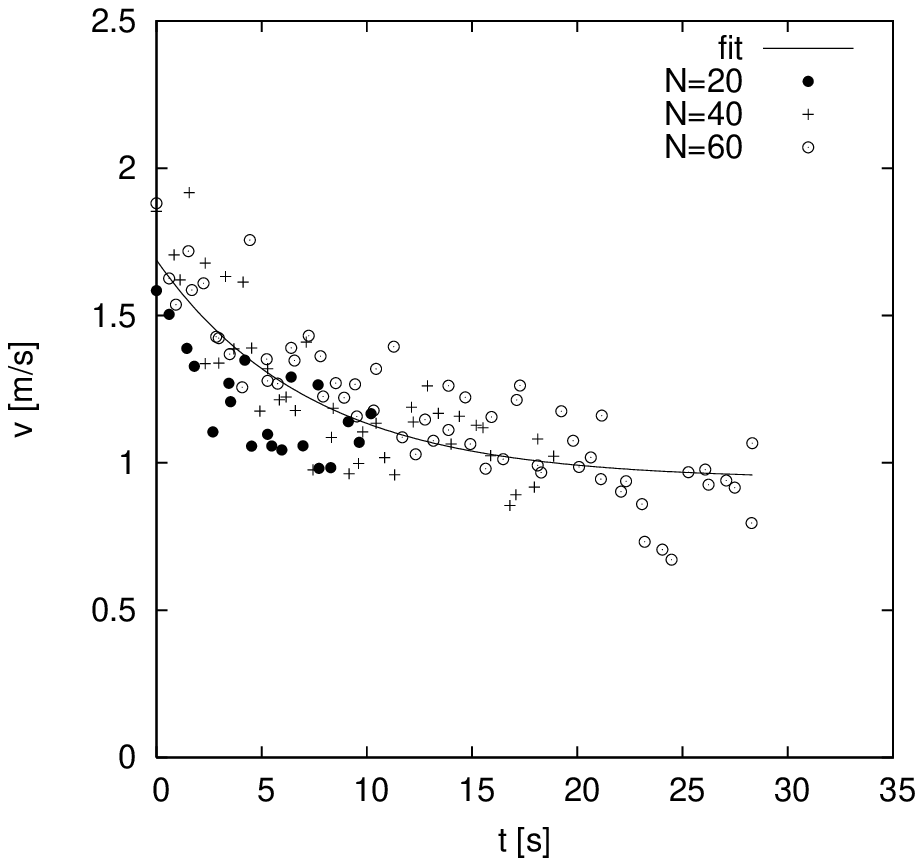}
\caption{Development of individual velocities in time for the run with $b=1.0 \; m$. 
The velocity decreases to a stationary value. For the runs with different $N$ the 
relaxation process is comparable.}
  \label{V-TIME-N}
  \end{center}
\end{figure}

The relaxation time $\tau$ characterizes the time in which a stationary state will be 
reached. The amplitude $A$ gives the difference between the stationary state and the 
initial velocity or density. The velocity or density at the stationary state is 
labeled $f_{stat}$. For the fit we use the data of all three runs for one width with 
different $N$, see Figure \ref{V-TIME-N}. The resulting stationary values and 
relaxation times are collected in Table \ref{FITRES}. Note, that the model function 
for the regression only describes the overall decrease in time and does not account 
for the density-fluctuations due to the small observation area or the fluctuations of 
the velocity in a stable state. Consequently we do not quote an error margin in 
Table \ref{FITRES}.
 
\begin{table}[h]
\begin{center}
\begin{tabular}{| c || c | c | c || c | c | c |}
\hline
\multicolumn{1}{| c ||}{} & \multicolumn{3}{| c ||}{\textbf{$f(t)=v(t)$}} & \multicolumn{3}{| c |}{\textbf{$f(t)=\rho(t)$}} \\
\hline
 $b\;[m]$ & $v_{stat}\; [m/s]$ & $A_{v} \; [m/s]$ & $\tau_{v}\; [s]$ & $\rho_{stat} \; [m^{-2}]$  & $A_{\rho} \; [m^{-2}]$ & $\tau_{\rho} \; [s]$ \\ \hline
\hline
0.8 & 1.18 & 0.354 & 3.55 & 1.42 & -1.82 & 0.24  \\ 
0.9 & 1.22 & 0.604 & 3.00 & 1.50 & -1.20 & 0.95  \\ 
1.0 & 1.17 & 0.485 & 3.83 & 1.59 & -1.87 & 0.31  \\ 
1.1 & 0.94 & 0.745 & 7.33 & 1.73 & -1.30 & 2.10  \\ 
1.2 & 0.99 & 0.836 & 5.63 & 1.70 & -1.28 & 1.45  \\ \hline
\end{tabular}
\end{center}
\caption[dummy]{Results for the fit to $v_i(t)$ and $\rho(t)$. \label{FITRES}}
\end{table}

The results of the regression analysis are collected in Table \ref{FITRES}. For 
$b \geq 1.0\;m$ even with $N=60$ the stationary state is not reached, see e.g. 
Figure \ref{V-R-TIME}. The results for $A$ and $\tau$ indicate that the relaxation 
into the stationary state is almost independent of the width. However, for a final 
judgment more data or a larger number of test persons would be necessary. 
Nevertheless, the results are accurate enough to check at which position of the 
fundamental diagram the stationary state will be located. Again, the increase of the 
stationary values for the density $\rho_{stat}$ can be explained by means of the 
zipper effect in combination with boundary effects. Given that the space at the 
boundaries can not be used as efficient as in the center of the bottleneck the density 
will increase for bigger width. While the stationary values of the densities increase 
the velocities shows a less pronounced decrease with $b$ according to the fundamental 
diagram. Again, for a final judgment an improved data basis is needed.

\section{Combined analysis with data from other experiments}
\label{RESULTS}

In this section we compare and combine our results with the data of previous 
measurements. We will show that the apparent discrepancies between the different flow 
measurements at bottlenecks can be traced back to different initial conditions. This 
allows a conclusive judgment on the question if the dependence of the flow and the 
bottleneck-width is linear. In addition does these results have a bearing on the 
question at which flow value a jam occurs.

\subsection{Comparison with the data of other experiments}

\begin{figure}[h]
  \begin{center}
  \includegraphics[width=.7\textwidth]{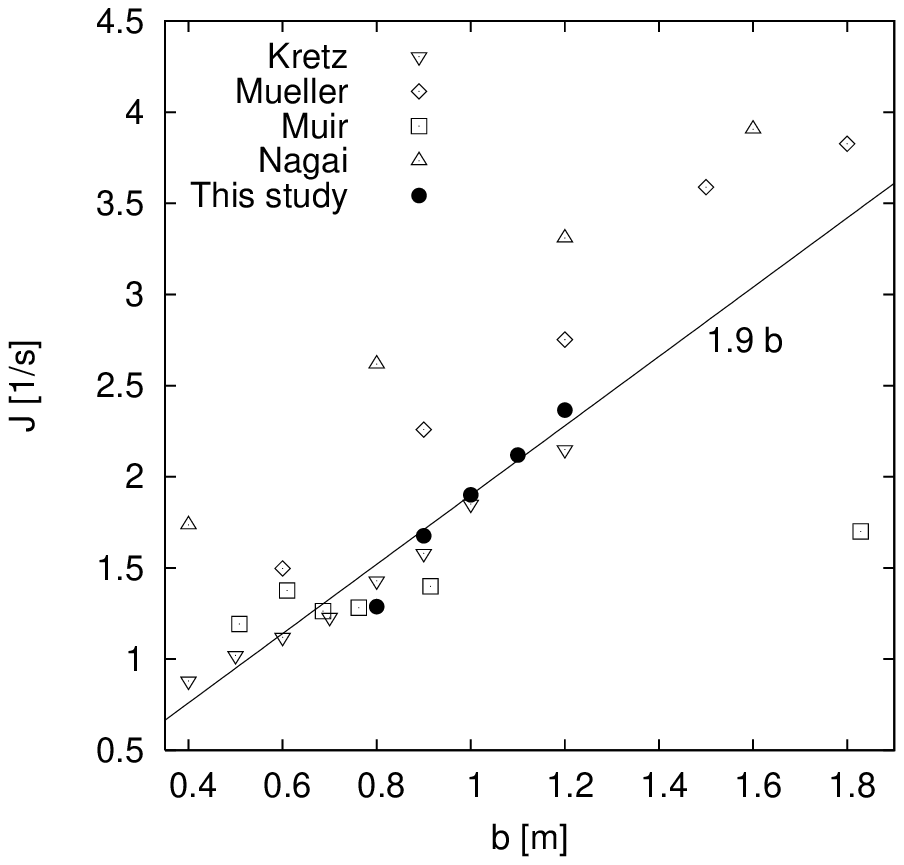}
  \caption{Influence of the width of a bottleneck on the flow. Experimental data from 
other authors at different types of bottlenecks and initial conditions in comparison
with the results of the above described experiment.}
  \label{PHI-B}
  \end{center}
\end{figure}

In Figure \ref{PHI-B} we have collected experimental data for flows through 
bottlenecks. All measurements were performed under laboratory conditions. The amount 
of test persons ranged from $N=30$ to $180$ persons. The influence of panic or pushing 
can be excluded as the collection is limited to measurements where the test persons 
were asked to move normally. 

\begin{figure}
  \centering
  \subfigure[Kretz]{\label{fig:suba}
    \includegraphics[width=.2\textwidth]{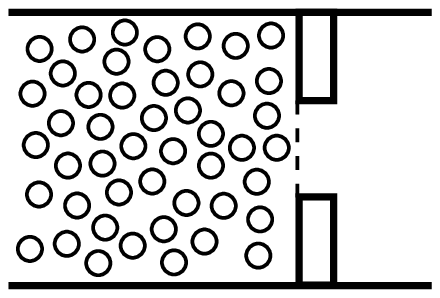}}
  \hspace{20.pt}
  \subfigure[Muir]{\label{fig:subb}
    \includegraphics[width=.2\textwidth]{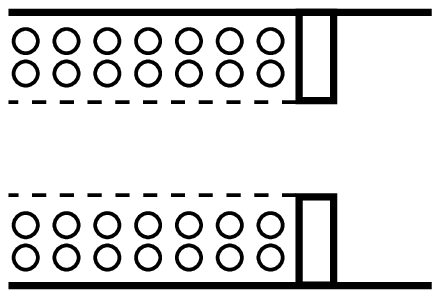}}
  \hspace{20.pt}
  \subfigure[M\"uller]{\label{fig:subc}
    \includegraphics[width=.2\textwidth]{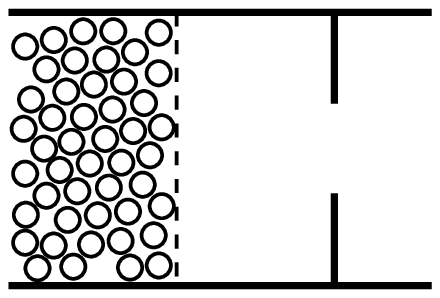}}
  \hspace{20.pt}
  \subfigure[Nagai]{\label{fig:subd}
    \includegraphics[width=.2\textwidth]{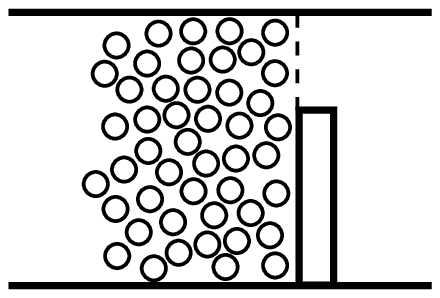}}
  \hspace{20.pt}
  \subfigure[This study]{\label{fig:sube}
    \includegraphics[width=.2\textwidth]{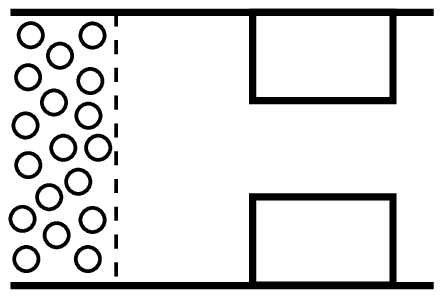}}
  \caption{Outlines of the experimental arrangements under which the data shown in Figure \ref{PHI-B} were taken.}
  \label{ESET:sub}
\end{figure} 

However, the experimental arrangements under which this data were taken differ in many 
detailed respects which provide possible explanations for the discrepancies. 
The outlines of the different experimental arrangements are sketched in Figure \ref{ESET:sub}. 
Significant differences concern:

\begin{itemize} 
\item the geometry of the bottleneck, i.e. its length and position with respect to the 
incoming flow.
\item the initial conditions, i.e. initial density values and the initial distance 
between the test persons and the bottleneck.  
\end{itemize}

The flow measurements of \cite{MUIR96} show a leveling off at $b>0.6\;m$. But the range 
of the flat profile from $b=0.6\;m$ to $b=1.8\;m$ indicates that obviously the passage 
width is not the limiting factor for the flow in this setup. 
The data of \cite{NAGAI06} and \cite{Mueller81} are shifted to higher 
flows in comparison with the data of \cite{KRETZ06c}, \cite{MUIR96} and our data. 
The height of the flows in the experiments of M\"uller and Nagai can be explained 
by their use of much higher initial densities which amount to $\rho_{ini} \approx 5\;m^{-2}$. 
That the initial density has this impact is confirmed by the study of Nagai et al., see 
Figure 6 in \cite{NAGAI06}. There it is shown that for $b=1.2\;m$ the flow grows from 
$J=1.04\; s^{-1}$ to $3.31\; s^{-1}$ when the initial density is increased from 
$\rho_{ini}=0.4\;m^{-2}$ to $5\; m^{-2}$. 

Next we will discuss the influence of the geometry and the position of the bottleneck. 
As can be seen in Figure \ref{PHI-B} there is a fair agreement between our data and 
the results obtained by Kretz. This indicates the minor importance of the 
bottleneck-length, given that in our experiment it amounts to $l_{bck}=2.8\;m$ while 
Kretz has chosen a bottleneck of $l_{bck}=0.4\;m$ length only. 

While in our and Kretz's experiment the bottleneck was centered, Nagai used a 
bottleneck located on the left side of the corridor. The comparison of the 
measurements for $b=1.2\;m$ of Nagai for the initial densities of $\rho=1.66\;m^{-2}$ 
(not displayed in Figure \ref{PHI-B}) with the data of Kretz and our results show that 
the flow of $J\approx 2.0\;s^{-1}$ is in good agreement with our measurements. The 
same applies to the flow measured by Muir which is particularly noteworthy since Muir 
has measured these values for the movement of a bottleneck in an airplane, where even 
the corridor is very small and the bottleneck is build by the galley units before the 
main embarkation point \cite{MUIR96}.

Finally we have to dicuss the differences between the flow values of Mueller and 
Nagai. As noted above this can not be explained by the initial densities since both
used similar (high) values. However, M\"uller measured at a simple opening in a thin 
wall of shelves while Nagai used a bottleneck with $l_{bck}=0.4\;m$. This fact could 
be used to explain higher flow values for M\"uller, given the common and 
reasonable assumption that the flow is correlated to the liberty of action which is 
higher for less deep bottlenecks. However, just the opposite is being observed, i.e. 
M\"uller's result lie below Nagai's data. A possible explanation is provided by the 
different distances from the bottleneck to the first pedestrians at time $t=0$. While 
Nagai positioned the test persons directly in front of the bottleneck M\"uller 
arranged a distance of $2.5\;m$ between the exit and the first pedestrians. Thus it is 
expected that the density will decrease during the movement from the initial position 
to the bottleneck resulting in an smaller density directly in front of the bottleneck. 

From all this we conclude that details of the bottleneck geometry and position play a 
minor role only while the initial density in front of the bottleneck has a major 
impact. 

\subsection{Linear dependence of flow and bottleneck-width \label{lin_dep_dis}}

As mentioned in the introduction one goal of this work is to examine if the flow or 
the capacity is a linear function of the width, $b$, of a bottleneck or if it grows in 
a stepwise manner, as suggested by \cite{Hoogendoorn05a}. Such a stepwise growth 
would question the validity of the specific flow concept used in most guidelines, see 
Section \ref{SEC-INTRO}.

However, the previous section has argued for the coherence of our data set and previous 
measurements. All of these results are compatible with a linear and continuous increase 
of the flow with the width of the bottleneck. Only around $b=0.7\;m$ it seems that the 
data of \cite{KRETZ06c} show a small edge. The edge is located exactly at the width where 
the zipper effect can begin to act, i.e. provides no evidence for a stepwise behavior in 
general.

Moreover does the alleged stepwise increase of the flow follows from the assumption 
that inside a bottleneck the formation of lanes with constant distance occurs.
In \cite{Hoogendoorn05a} this assumption is based on flow measurements at two different
bottlenecks at $b=1\;m$ and $b=2\;m$. It is doubtful whether this results can be 
extrapolated to intermediate values of the width. In fact our data show no evidence 
for the appearance of lanes with constant distance (see Section 
\ref{hoogi_macht_mist}, in particular Figure \ref{MULTITRJ}).

\subsection{The criteria for jam occurrence}

The above results can be used to address a crucial question in pedestrian dynamics, 
namely the criteria for the occurrence of a jam. As mentioned already in the 
introduction it is commonly assumed that this happens when the incoming flow (through 
a surface in front of the bottleneck) exceeds the capacity of the bottleneck. 
Here the capacity of the bottleneck is defined as the maximum of the specific 
flow, $J_s(\rho)$, times its width. 

A preliminary analysis of the situation one meter in front of the bottleneck 
entrance shows density fluctuations inside the jam between $\rho=4$ to $6\;m^{-2}$ 
independent from the width of the bottleneck.
Our results from Section \ref{ANA} can be used to examine which density and 
flow inside the bottleneck is present for a situation where a jam occurs in 
front of the bottleneck. 
For this purpose the fitted values for $\rho_{stat}$ and $v_{stat}$ are used 
(see Table \ref{FITRES}).

\begin{figure}[htb]
        \begin{center}
  \includegraphics[width=.7\textwidth]{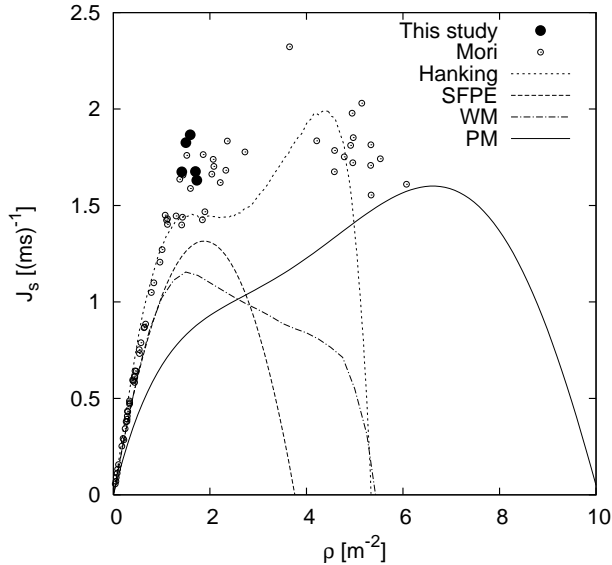}
  \caption{Experimental data of the flow and the associated density in the bottleneck 
in comparison with experimental data for the fundamental diagram of unidirectional 
pedestrian streams and the specifications for the fundamental diagram according to the 
SFPE Handbook and the guidelines of Weidmann and Predtechenskii and Milinskii.}
  \label{R-FS}
  \end{center}
\end{figure}

Figure \ref{R-FS} indicates that the values for the stationary density derived from 
our experiment are exactly located at the position where the fundamental diagram 
according to the SFPE-Handbook and the guideline of Weidmann show the maximum of the 
flow (while the absolute value of the flow exceeds the predicted values), i.e.  
$\rho \approx 1.8\; m^{-2}$. This seems to support the common jam-occurrence criteria. 
However, two observations cast doubt on this conclusion. Already when discussing the 
data of M\"uller and Nagai we have mentioned that higher initial densities result in 
higher flow values, i.e. that the maximal flow can not be near $\rho = 1.8\; m^{-2}$. 
In addition do the fundamental diagrams of Mori \cite{MORI_87}, Hanking \cite{HANK58} 
and PM \cite{PRED78ENG} display a completely different shape. According to Mori and 
Hanking and in agreement with the specification of PM the flow will increase from 
$\rho \approx 1.8\; m^{-2}$ or stay constant with increasing density. Moreover does 
the level of the flow measured in our experiment conforms much better with their 
specification. 

In the literature one finds a multitude of different explanations for the deviations 
between fundamental diagrams, including cultural and population differences 
\cite{helbing07a}, differences between uni- and multidirectional flow 
\cite{NAV69,PUSH75,LAM03}, influence of psychological factors given by the incentive 
of the movement \cite{PRED78ENG} and, partially related to the latter, the type of 
traffic like commuters or shoppers \cite{OEDI63}. Moreover, in a laboratory 
experiment, the population is special, and the incentive will most likely be 
different from that in a field measurements. We assume that for densities higher than 
$\rho=1.8\;m^{-2}$ cultural and population differences influence the required space 
and change the scaling of the density but not the shape of the diagram. The change 
in the scaling of the fundamental diagram due to the incentive of the movement and 
body size is accordance with Predtechenskii and Milinskii \cite{PRED78ENG}. As a 
starting point of a further discussion we refer to influences due to multidirectional 
flow \cite{NAV69,PUSH75,LAM03} and look in the origin of the fundamental diagrams 
shown in Figure~\ref{R-FS}. It seems that the most detailed one is given by Weidmann 
which used 25 data sets for generating his fundamental diagram. But if one performs an 
examination of the data which went into Weidmann's combination it turns out 
that most measurements with densities larger then $\rho=1.8\; m^{-2}$ are performed 
with bidirectional streams, see \cite{OLD68,NAV69,OEDI63,OFL72,POL83}. 
But also data gained by measurements on strictly unidirectional streams has been 
considered \cite{HANK58,FRUIN71}. The data of Hanking and Wright gained by 
measurements in London subway (UK) are in good agreement with the data of Mori and 
Tsukaguchi who measured in the central business district of Osaka (Japan).
Their fundamental diagrams agree very well and differ notable from the fundamental 
diagram of the SFPE and Weidmann for densities larger $\rho=1.8\;m^{-2}$. In 
accordance with Predtechenskii and Milinskii we assume that the maximum of the 
fundamental diagram for unidirectional movement is at much higher densities. Thus 
we conclude from our measurements of densities and velocities that the flow through 
the bottleneck under normal conditions will not reach the capacity defined through 
the maximum of the fundamental diagram and that a jam occurs well before the 
maximum is reached.

There are three possible reasons for jamming below the capacity limit. One is clearly 
flow fluctuations. Local density maxima higher then capacity-density cause obstructions 
that can be resolved only if the incoming flow is less than the outflow, which in this 
case is below capacity. A second reason is in the local organization of the 
pedestrians. At high densities, the full flow requires a fairly regular positioning of 
persons. This may be achieved by pre-arranging them - as in the experiments in 
\cite{Mueller81,NAGAI06} - or possibly by funneling the flow into the bottleneck. 
Neither the random arrival at a narrow passage out of the almost free motion in a wide 
room nor the - also random - starting into the passage from waiting in a jam can 
achieve this. Therefor the actual bottleneck is the entrance into the passage, while 
in the passage itself the flow corresponds to the lower density associated with the 
observed flux (compare PM \cite{PRED78ENG}). The third reason is motivational, people 
usually prefer larger distances to the person in front than necessary unless there is 
the danger of somebody else filling the gap - or some other incentive for moving up 
close. These topics will be the subject of future research.     

\section{Summary}

We have studied experimentally the flow of unidirectional pedestrian streams through 
bottlenecks under normal conditions. For the data analysis we used the trajectories to 
determine the time development of the individual velocities, local densities and time 
gaps for different widths of the bottleneck. The analysis shows that for a small 
variation of the width quantities like the time gap distribution or the lane distance 
change continuously if the zipper effect is acting. The results for velocities and 
densities are compared to common methods for the capacity estimation and to 
experimental data from other studies.

Except for the edge at $b\approx0.7\;m$ due to zipper effect starting to act all 
collected data for flow measurements are compatible with a linear and continuous 
increase with the bottleneck-width. The linear dependency between the flow and 
the width holds for different kinds of bottlenecks and initial conditions. Hence 
the basic flow equation in combination with the use of the specific flow concept 
is justified for facilities with $b>0.7\;m$. Moreover does the comparison with 
flow measurements through bottlenecks of different types and lengths show that 
the exact geometry of the bottleneck is of only minor influence on the flow. 

The rise of the flow through the bottleneck with an increase of the initial density 
in front of the bottleneck from $\rho=1.8\;m^{-2}$ to $5 \; m^{-2}$ indicates at 
least for this density region a rising of the slope of the fundamental diagram for 
unidirectional streams. This is in disagreement with most fundamental diagrams as 
documented in handbooks and guidelines but in agreement with empirical fundamental 
diagrams restricted to unidirectional movement and with the shape of the fundamental 
diagram according to Predtechenskii and Milinskii. Thus the capacity defined by the 
maximum of the fundamental diagram is expected for densities substantial higher than 
$\rho=1.8\;m^{-2}$. With respect to the absolute flow values one needs to bear in mind 
that our results were obtained with young probands and under laboratory conditions.

Our measurements of densities and velocities inside the bottleneck show that the 
stationary flow will tune around densities of $\rho \approx 1.8\; m^{-2}$ when a 
jam occurs in front of the bottleneck. 
For higher densities higher flow values are expected in reference to the maxima of 
fundamental diagrams restricted to unidirectional streams. Supplementary an increase 
of the flow for $5.0\; m^{-2}>\rho>1.8\; m^{-2}$ is observed in the experiments of 
\cite{NAGAI06}. Thus jamming occurs for flow values lower than maximal possible 
flow values.
This contradicts the common assumption that a jam appears only when the capacity is 
exceeded. While the continuity equation trivially 
implies that the capacity is the upper limit for a jam producing flow, do our results 
indicate that this happens for smaller flow values already. Possible reasons are 
stochastic flow fluctuation, flow interferences due to the necessity of local 
organization or psychological founded changes of the incentive during the access into 
the bottleneck.  

All experimental measurements of flow values through bottlenecks discussed in this 
paper are gained by laboratory experiments. Concerning differences between field 
measurements and laboratory experiments one has to note, that absolute values and 
figures obtained in a laboratory experiment cannot simply be used for design 
recommendations. The qualitative conclusions however - the bottleneck capacity grows 
about linear with width and jamming may occur below maximum capacity - are not affected 
by changes in scaling of the flow measurements or of the fundamental diagram.

Tragic disasters like at the L\"owenbr\"au-Keller in Munich, Germany (1973), Bergisel 
in Austria (1999) or Akashi Japan (2001) \cite{TSUJI03} have shown that already 
without the influence of panic insufficient dimensions of facilities can cause high 
densities in bottlenecks. At high densities small interferences in the crowd can not 
be balanced and may cause fatalities due to people stamped to the ground by high 
pressure. For a better evaluation of pedestrian facilities with respect to safety 
it is necessary to explore under which circumstances such an increase of the density 
inside the bottleneck occurs.

As a first step our results can be used to improve the capacity estimations in
planing guidelines and handbooks to prevent such situations.\\ 

{\bf Acknowledgments}

We thank Tinh Nguyen for the extraction of the literature data for the fundamental 
diagrams. For the useful discussions we thank Tobias Kretz, Christian Rogsch and 
Andreas Schadschneider. For the informations and support concerning experimental data 
published in \cite{NAGAI06} we thank Takashi Nagatani.


\begin{thebibliography}{10}

\bibitem{ADOBE}
Adobe.
\newblock {Adobe After Effects}, 2005.

\bibitem{CAR70}
R.~L. Carstens and S.~L. Ring.
\newblock Pedestrian capacities of shelter entrances.
\newblock {\em Traffic Engineering}, 41:38--43, 1970.

\bibitem{SFPE02}
P.~J. DiNenno.
\newblock {\em SFPE Handbook of Fire Protection Engineering}.
\newblock National Fire Protection Association, Quincy MA, third edition, 2002.

\bibitem{FRUIN71}
J.~J. Fruin.
\newblock {\em {Pedestrian Planning and Design}}.
\newblock Elevator World, New York, 1971.

\bibitem{HANK58}
B.~D. Hankin and R.~A. Wright.
\newblock Passenger flow in subways.
\newblock {\em Operational Research Quarterly}, 9:81--88, 1958.

\bibitem{HCPM85}
{\em Highway Capacity Manual}.
\newblock Washington DC, 1985.

\bibitem{helbing06a}
D.~Helbing, A.~Johansson, J.~Mathiesen, M.H. Jensen, and A.~Hansen.
\newblock Analytical approach to continuous and intermittent bottleneck flows.
\newblock {\em Physical Review Letters}, 97:168001, 2006.

\bibitem{helbing07a}
Dirk Helbing, Anders Johansson, and Habib~Zein Al-Abideen.
\newblock The dynamics of crowd disasters: An empirical study.
\newblock {\em Physical Review E}, 75:046109, 2007.

\bibitem{Hoogendoorn05a}
S.~P. Hoogendoorn and W.~Daamen.
\newblock Pedestrian behavior at bottlenecks.
\newblock {\em Transportation Science}, 39 2:0147--0159, 2005.

\bibitem{Hoogendoorn03a}
S.~P. Hoogendoorn, W.~Daamen, and P.~H.~L. Bovy.
\newblock Microscopic pedestrian traffic data collection and analysis by
  walking experiments: Behaviour at bottlenecks.
\newblock In E.~R. Galea, editor, {\em Pedestrian and Evacuation Dynamics '03},
  pages 89--100. CMS Press, London, 2003.

\bibitem{MINUIT}
F.~James.
\newblock {MINUIT - Function Minimization and Error Analysis}, 1994.
\newblock CERN Program Library entry D506.

\bibitem{KRETZ06c}
T.~Kretz, A.~Gr{\"u}nebohm, and M.~Schreckenberg.
\newblock Experimental study of pedestrian flow through a bottleneck.
\newblock {\em J. Stat. Mech.}, page P10014, 2006.

\bibitem{LAM03}
W.~H.~K. Lam, J.Y.S. Lee, and C.~Y. Cheung.
\newblock {A generalised function for modeling bi-directional flow effects in
  indoor walkways in Hong Kong}.
\newblock {\em Transportation Research Part A}, 37:789--810, 2003.

\bibitem{MINT51}
A.~Mintz.
\newblock Non-adaptive group behaviour.
\newblock {\em The Journal of abnormal and social psychology}, 46:150--159,
  1951.

\bibitem{MORI_87}
M.~Mori and H.~Tsukaguchi.
\newblock A new method for evaluation of level of service in pedestrian
  facilities.
\newblock {\em Transp. Res. Part A}, 21A(3):223--234, 1987.

\bibitem{MUIR96}
H.~C. Muir, D.~M. Bottomley, and C.~Marrison.
\newblock Effects of motivation and cabin configuration on emergency aircraft
  evacuation behavior and rates of egress.
\newblock {\em The International Journal of Aviation Psychology}, 6(1):57--77,
  1996.

\bibitem{Mueller81}
K.~M{\"u}ller.
\newblock {\em {Die Gestaltung und Bemessung von Fluchtwegen f{\"u}r die
  Evakuierung von Personen aus Geb{\"a}uden}}.
\newblock dissertation, {Technische Hochschule Magdeburg}, 1981.

\bibitem{NAGAI06}
R.~Nagai, M.~Fukamachi, and T.~Nagatani.
\newblock Evacuation of crawlers and walkers from corridor through an exit.
\newblock {\em Physica A}, 367:449--460, 2006.

\bibitem{NAV69}
P.~D. Navin and R.~J. Wheeler.
\newblock Pedestrian flow characteristics.
\newblock {\em Traffic Engineering}, 39:31--36, 1969.

\bibitem{NELS02}
H.~E. Nelson and F.~W. Mowrer.
\newblock Emergency movement.
\newblock In P.~J. DiNenno, editor, {\em SFPE Handbook of Fire Protection
  Engineering}, chapter~14, page 367. National Fire Protection Association,
  Quincy MA, third edition, 2002.

\bibitem{OEDI63}
D.~Oeding.
\newblock {Verkehrsbelastung und Dimensionierung von Gehwegen und anderen
  Anlagen des Fu{\ss}g{\"a}ngerverkehrs}.
\newblock Forschungsbericht~22, {Technische Hochschule Braunschweig}, 1963.

\bibitem{OFL72}
C.~A. O'Flaherty and M.~H. Parkinson.
\newblock Movement in a city centre footway.
\newblock {\em Traffic Engineering and Control}, pages 434--438, Feb. 1972.

\bibitem{OLD68}
S.~J. Older.
\newblock Movement of pedestrians on footways in shopping streets.
\newblock {\em Traffic Engineering and Control}, 10:160--163, 1968.

\bibitem{POL83}
A.~Polus, J.~L. Joseph, and A.~Ushpiz.
\newblock Pedestrian flow and level of service.
\newblock {\em Journal of Transportation Engineering}, 109 1:46--56, 1983.

\bibitem{PRED78ENG}
V.~M. Predtechenskii and A.~I. Milinskii.
\newblock {\em Planing for foot traffic flow in buildings}.
\newblock Amerind Publishing, New Dehli, 1978.
\newblock Translation of: Proekttirovanie Zhdanii s Uchetom Organizatsii
  Dvizheniya Lyuddskikh Potokov, Stroiizdat Publishers, Moscow, 1969.

\bibitem{PUSH75}
B.~Pushkarev and J.~M. Zupan.
\newblock Capacity of walkways.
\newblock {\em Transportation Research Record}, 538:1--15, 1975.

\bibitem{RUPDIP06}
T.~Rupprecht.
\newblock {Untersuchung zur Erfassung der Basisdaten von Personenstr{\"o}men}.
\newblock diplomathesis, Bergische Universit{\"a}t Wuppertal, FB-D Abt.
  Sicherheitstechnik, 2006.

\bibitem{TSUJI03}
Yutaka Tsuji.
\newblock Numerical simulation of pedestrian flow at high densities.
\newblock In E.~R. Galea, editor, {\em Pedestrian and Evacuation Dynamics '03},
  pages 27--38. CMS Press, London, 2003.

\bibitem{UN05}
United Nations. 2005
\newblock http://esa.un.org/unup/p2k0data.asp.

\bibitem{WEID93}
U.~Weidmann.
\newblock {Transporttechnik der Fu{\ss}g{\"a}nger}.
\newblock {Schriftenreihe des IVT}~90, {ETH Z{\"u}rich}, 1993.

\end{thebibliography}


\end{document}